\documentclass{iau}
\usepackage{times}
\usepackage{graphicx}
\usepackage{natbib}
\usepackage{hyperref}
\hypersetup{
    final=true,
    pageanchor=true,
    colorlinks=true,
    breaklinks=true,
    linkcolor=blue,
    citecolor=blue,
    urlcolor=blue,
    pdfpagemode=UseNone,
    pdftitle={Digitization of Spoerer's sunspot drawings},
    pdfauthor={A. Diercke, R. Arlt, &  C. Denker},
    pdfsubject={Solar Physics}}
\usepackage{nicefrac}
\usepackage{amssymb}
\usepackage[usenames, dvipsnames]{color}
\usepackage[modulo,switch]{lineno}
\newcommand{\arcdeg}{\mbox{$^{\circ}$}}

\title{Digitization of Sp\"orer's sunspot drawings}

\author[A.~Diercke, R.~Arlt, \&  C.~Denker]   %% give here short author list %%
{Andrea Diercke$^{1,2}$,
 Rainer Arlt$^1$, \and Carsten Denker$^1$}

\affiliation{$^1$ Leibniz-Institut f\"ur Astrophysik Potsdam,
    14482 Potsdam, Germany \\ email: \textcolor{blue}{{\tt adiercke@aip.de}} \\[\affilskip]
$^2$ Institut f\"ur Physik und Astronomie,
Universit\"at Potsdam,
14476 Potsdam,
Germany }

\pubyear{2012}
\volume{294}  %% insert here IAU Symposium No.
\pagerange{119--126}
\jname{Solar and Astrophysical Dynamos and Magnetic Activity}
\editors{A.G.~Kosovichev, E.M.~de Gouveia Dal Pino \&  Y.~Yan, eds.}
\begin{document}

\maketitle

\begin{abstract}
 Much of our knowledge about the solar dynamo is based on sunspot observations. It is thus desirable to extend the set of positional and morphological data of sunspots into the past. Gustav Sp\"orer observed in Germany from Anklam (1861--1873) and Potsdam (1874--1894). He left detailed prints of sunspot groups, which we digitized and processed to mitigate artifacts left in the print by the passage of time. After careful geometrical correction, the sunspot data are now available as synoptic charts for almost 450 solar rotation periods. Individual sunspot positions can thus be precisely determined and spot areas can be accurately measured using morphological image processing techniques. These methods also allow us to determine tilt angles of active regions (Joy's law) and to assess the complexity of an active region.
\end{abstract}

\section{Sunspot observations and digitization of sunspot drawings}
The first continuous observations of sunspots were made by Samuel Heinrich Schwabe 1825--1867. In this time, Richard Carrington started his observations in 1853, followed by Sp\"orer's first observations in January 1861 during Carrington rotation period No.~96. In total 445 rotation periods were recorded. The images were published as lithographs, which reproduce many details of the sunspot fine structure, e.g., umbra and penumbra \citep{Spoerer1874}.
Each page contains five rotation periods. The abscissa shows the heliographic longitude $L $ from $360\arcdeg$ to $0\arcdeg$ and on the ordinate is the heliographic latitude $b$ from $+40\arcdeg$ to $-40\arcdeg$ for one rotation period.

The sunspot drawings were collated in hardbound books. Consequently, pages will not be perfectly flat when opening the books. Binder clips and metal coins were used to get a flat image of the drawings. Whole pages were photographed with a digital camera. The drawings were published nearly 150 years ago, thus, the pages are yellowish, stained, and corrugated. These artifacts have to be removed because they would severely affect the analysis of sunspot properties. All algorithms were developed in the Interactive Data Language (IDL).
\begin{figure*}[t]
\centering
\includegraphics[width=0.94\textwidth]{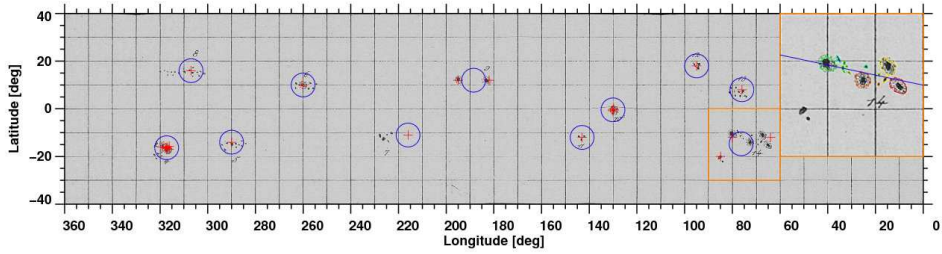}
\caption{Image after remapping, where crosses mark the sunspot positions and circles the mean positions of the sunspot groups.% as listed by Sp\"orer's sunspot tables.
 The positions were measured over several days, i.e., it is possible to visualize the disk passage of sunspot groups. Sometimes there are errors in the data, e.g., the table entry of spot 7 differs from its position in the image. The insert shows the results of blob analysis for active region 14.}
\label{Fig:map}
\end{figure*}

\section{Image processing and sunspot morphology}
After converting the color images to gray-scale images, they are rotated by $90\arcdeg$ to have them in portrait format as in the books. The images are cropped so that only the coordinate grids with sunspot data and their annotation remain. To remove a small rotation angle error of the camera, the standard deviation of the intensity traces of all rows and all columns has to reach a maximum. Then, the angular offset is zero.
% Therefore, we rotated the images in steps of $0.1\arcdeg$ within the range of $\pm1\arcdeg$ and computed the respective standard deviations. A parabolic fit to this curve yields the exact angle by which the images have to be rotated.
 An order-statistics filter \citep{Gonzalez2002} is used to compute a background intensity map. The filter ranks the intensities within a $33 \times 33$ pixel neighborhood and replaces the value of the center pixel by the $80^{\mathrm{th}}$ percentile of the intensity distribution. The 16-pixel wide borders of the full background intensity map are replaced by the mean background intensity. Division of an image by its background intensity map (i.e., a pseudo flat-fielding procedure) results in an artificially bleached and flattened image. 

To get the exact coordinates of the grid lines, a $17 \times 17$-pixel kernel is used, where the value of two intersections will be set to unity and all other pixels at zero. A good approximation of the grid coordinates is the average of the line profiles of rows and columns, which were smoothed with a $17$-pixel Lee-filter and afterwards normalized. Good estimates of the $x$- and $y$-grid coordinates yields a parabolic fit to local maxima of row and column profile. To improve the grid coordinates, from the binary mask, a region of $20$ pixels will be extracted and enlarged by factor of five. Afterwards, the region will be smoothed with a $5$-pixel boxcar filter. Nevertheless, sunspots near to intersecting grid lines can lead to erroneous results. As a precaution, all $x$- and $y$-coordinates are checked against the positions of their immediate neighbors and replaced, if a significant offset is detected.
The heliographic longitude and latitude are given in increments of $10\arcdeg$, so that we have a grid of ($37 \times 9$) $x$- and $y$-coordinates for each rotation period. We also included a $5\arcdeg$-wide boundary to minimize sampling errors at the periphery. Subsequently, Delaunay triangulations were used to interpolate the sunspot data to the new grid (the result is shown in Fig.~\ref{Fig:map}). %Data of this type ($3600 \times 800$ pixels) are obtained for each of the 445 rotation periods. Finally, we use morphological image processing to remove the grid lines.

After image processing, standardized sunspot maps are at hand for further analysis. The sunspot drawings are made of dark-inked areas for the umbrae and many single points to indicate the penumbral fine structure. A more realistic `intensity distribution' can be achieved using some Gaussian smoothing. Even though blurred, the images are now suitable for image processing techniques such as intensity thresholding. Morphological opening and closing operations can then be applied to the binary masks, in which individual sunspots are labeled. Thus, `holes' in the sunspot area can be filled and spot contours become less rough. We used standard tools for `blob analysis' \citep{Fanning2011} to derive parameters describing the morphology of sunspots and to determine their positions. In general, this approach leads to very good results but in some cases, the spoke-like structure used to indicate penumbral fine structure results in artifacts. Another problem are the numbers, which label the sunspot groups. However, since they consist of thin lines as compared to the more compact sunspots, morphological image processing is again capable of indentifying and removing the labels. In summary, Sp\"orer's sunspot drawing are of high quality facilitating the application of modern feature recognition and image processing methods.

\acknowledgements AD and CD were supported by grant DE~787/3-1 of the Deutsche
Forschungsgemeinschaft (DFG). The authors thank the AIP librarian Regina von Berlepsch for her extended support.


\begin{thebibliography}{3}
\bibitem[\protect\citeauthoryear{Sp\"orer}{1874}]{Spoerer1874}
    Sp\"orer, G.: 1874, Pub.\ Astron.\ Ges.\ 13
\bibitem[\protect\citeauthoryear{Fanning}{2011}]{Fanning2011}
   Fanning, D.W.: 2011, \textit{Coyote's Guide to Traditional IDL Graphics},
    Coyote Book Publ., Fort Collins, CO
\bibitem[\protect\citeauthoryear{Gonzalez \& Woods}{2002}]{Gonzalez2002}
    Gonzalez, R.C., Woods, R.E.: 2002, \textit{Digital Image Processing},
    Prentice-Hall, Upper Saddle River, NJ
\end{thebibliography}
\end{document}